\documentclass[dvips]{article}
\usepackage{supertabular,lscape,epsfig}
\usepackage{amssymb}
\usepackage{amsmath}
\DeclareSymbolFont{ppa}{OT1}{ppl}{m}{it}
\DeclareMathSymbol{\vv}{\mathalpha}{ppa}{'166}

\begin{document}

\newcommand{\dd}{\,{\rm d}}
\newcommand{\ie}{{\it i.e.},\,}
\newcommand{\etal}{{\it et al.\ }}
\newcommand{\eg}{{\it e.g.},\,}
\newcommand{\cf}{{\it cf.\ }}
\newcommand{\vs}{{\it vs.\ }}
\newcommand{\zdot}{\makebox[0pt][l]{.}}
\newcommand{\up}[1]{\ifmmode^{\rm #1}\else$^{\rm #1}$\fi}
\newcommand{\dn}[1]{\ifmmode_{\rm #1}\else$_{\rm #1}$\fi}
\newcommand{\upd}{\up{d}}
\newcommand{\uph}{\up{h}}
\newcommand{\upm}{\up{m}}
\newcommand{\ups}{\up{s}}
\newcommand{\arcd}{\ifmmode^{\circ}\else$^{\circ}$\fi}
\newcommand{\arcm}{\ifmmode{'}\else$'$\fi}
\newcommand{\arcs}{\ifmmode{''}\else$''$\fi}
\newcommand{\MS}{{\rm M}\ifmmode_{\odot}\else$_{\odot}$\fi}
\newcommand{\RS}{{\rm R}\ifmmode_{\odot}\else$_{\odot}$\fi}
\newcommand{\LS}{{\rm L}\ifmmode_{\odot}\else$_{\odot}$\fi}

\newcommand{\Abstract}[2]{{\footnotesize\begin{center}ABSTRACT\end{center}
\vspace{1mm}\par#1\par
\noindent
{~}{\it #2}}}

\newcommand{\TabCap}[2]{\begin{center}\parbox[t]{#1}{\begin{center}
  \small {\spaceskip 2pt plus 1pt minus 1pt T a b l e}   
  \refstepcounter{table}\thetable \\[2mm]
  \footnotesize #2 \end{center}}\end{center}}

\newcommand{\TableSep}[2]{\begin{table}[p]\vspace{#1}
\TabCap{#2}\end{table}}

\newcommand{\FigCap}[1]{\footnotesize\par\noindent Fig.\  %
  \refstepcounter{figure}\thefigure. #1\par}

\newcommand{\TableFont}{\footnotesize}
\newcommand{\TableFontIt}{\ttit}
\newcommand{\SetTableFont}[1]{\renewcommand{\TableFont}{#1}}

\newcommand{\MakeTable}[4]{\begin{table}[p]\TabCap{#2}{#3}
  \begin{center} \TableFont \begin{tabular}{#1} #4 
  \end{tabular}\end{center}\end{table}}

\newcommand{\MakeTableTop}[4]{\begin{table}[t]\TabCap{#2}{#3}
  \begin{center} \TableFont \begin{tabular}{#1} #4 
  \end{tabular}\end{center}\end{table}}

\newcommand{\MakeTableSep}[4]{\begin{table}[p]\TabCap{#2}{#3}
  \begin{center} \TableFont \begin{tabular}{#1} #4
  \end{tabular}\end{center}\end{table}}
\newcommand{\TabCapp}[2]{\begin{center}\parbox[t]{#1}{\centerline{
  \small {\spaceskip 2pt plus 1pt minus 1pt T a b l e}
  \refstepcounter{table}\thetable}
  \vskip2mm
  \centerline{\footnotesize #2}}
  \vskip3mm
\end{center}}

\newcommand{\MakeTableSepp}[4]{\begin{table}[p]\TabCapp{#2}{#3}\vspace*{-.7cm}
  \begin{center} \TableFont \begin{tabular}{#1} #4 
  \end{tabular}\end{center}\end{table}}

\newfont{\bb}{ptmbi8t at 12pt}
\newfont{\bbb}{cmbxti10}
\newfont{\bbbb}{cmbxti10 at 9pt}
\newcommand{\uprule}{\rule{0pt}{2.5ex}}
\newcommand{\douprule}{\rule[-2ex]{0pt}{4.5ex}}
\newcommand{\dorule}{\rule[-2ex]{0pt}{2ex}}
\def\thefootnote{\fnsymbol{footnote}}

\newenvironment{references}%
{
\footnotesize \frenchspacing
\renewcommand{\thesection}{}
\renewcommand{\in}{{\rm in }}
\renewcommand{\AA}{Astron.\ Astrophys.}
\newcommand{\AAS}{Astron.~Astrophys.~Suppl.~Ser.}
\newcommand{\ApJ}{Astrophys.\ J.}
\newcommand{\ApJS}{Astrophys.\ J.~Suppl.~Ser.}
\newcommand{\ApJL}{Astrophys.\ J.~Letters}
\newcommand{\AJ}{Astron.\ J.}
\newcommand{\IBVS}{IBVS}
\newcommand{\PASP}{P.A.S.P.}
\newcommand{\Acta}{Acta Astron.}
\newcommand{\MNRAS}{MNRAS}
\renewcommand{\and}{{\rm and }}
\section{{\rm REFERENCES}}
\sloppy \hyphenpenalty10000
\begin{list}{}{\leftmargin1cm\listparindent-1cm
\itemindent\listparindent\parsep0pt\itemsep0pt}}%
{\end{list}\vspace{2mm}}

\def\TYLDA{~}
\newlength{\DW}
\settowidth{\DW}{0}
\newcommand{\dw}{\hspace{\DW}}

\newcommand{\refitem}[5]{\item[]{#1} #2%
\def\REFARG{#3}\ifx\REFARG\TYLDA\else, {\it#3}\fi
\def\REFARG{#4}\ifx\REFARG\TYLDA\else, {\bf#4}\fi
\def\REFARG{#5}\ifx\REFARG\TYLDA\else, {#5}\fi.}

\newcommand{\Section}[1]{\section{\hskip-6mm.\hskip3mm#1}}
\newcommand{\Subsection}[1]{\subsection{#1}}
\newcommand{\Acknow}[1]{\par\vspace{5mm}{\bf Acknowledgements.} #1}
\pagestyle{myheadings}

\newcommand{\xrule}{\rule{0pt}{2.5ex}}
\newcommand{\xxrule}{\rule[-1.8ex]{0pt}{4.5ex}}
\def\thefootnote{\fnsymbol{footnote}}
\hyphenation{OSARGs}

\begin{center}
{\Large\bf The Optical Gravitational Lensing Experiment.\\
\vskip3pt
Period--Luminosity Relations\\
\vskip6pt
of Variable Red Giant Stars\footnote{Based on
observations obtained with the 1.3-m Warsaw telescope at the Las Campanas
Observatory of the Carnegie Institution of Washington.}}
\vskip1.0cm
{\bf I.~~S~o~s~z~y~\'n~s~k~i$^1$,~~W.\,A.~~D~z~i~e~m~b~o~w~s~k~i$^1$,
~~A.~~U~d~a~l~s~k~i$^1$,\\
M.~~K~u~b~i~a~k$^1$,~~M.\,K.~~S~z~y~m~a~{\'n}~s~k~i$^1$,~~G.~~P~i~e~t~r~z~y~\'n~s~k~i$^{1,2}$,\\
\L.~~W~y~r~z~y~k~o~w~s~k~i$^{1,3}$,~~O.~~S~z~e~w~c~z~y~k$^{1,2}$,~~and~~K.~~U~l~a~c~z~y~k$^1$}
\vskip8mm
{$^1$Warsaw University Observatory, Al.~Ujazdowskie~4, 00-478~Warszawa,
Poland\\
{\small e-mail: 
(soszynsk,wd,udalski,mk,msz,pietrzyn,wyrzykow,szewczyk,kulaczyk)@astrouw.edu.pl}\\
$^2$ Universidad de Concepci{\'o}n, Departamento de Fisica, Casilla 160--C,
Concepci{\'o}n, Chile\\ 
$^3$ Institute of Astronomy, University of Cambridge, Madingley Road,
Cambridge CB3 0HA, UK}
\end{center}

\vskip1.2cm

\Abstract{Period--luminosity (PL) relations of variable red giants in the
Large (LMC) and Small Magellanic Clouds (SMC) are presented. The PL
diagrams are plotted in three planes: $\log P$--$K_S$, $\log P$--$W_{JK}$,
and $\log P$--$W_I$, where $W_{JK}$ and $W_I$ are reddening free Wesenheit
indices. Fourteen {\it PL} sequences are distinguishable, and some of them
consist of three closely spaced ridges. Each of the sequences is fitted
with a linear or quadratic function. The similarities and differences
between the {\it PL} relations in both galaxies are discussed for four
types of red giant variability: OGLE Small Amplitude Red Giants (OSARGs),
Miras and Semiregular Variables (SRVs), Long Secondary Periods (LSPs) and
ellipsoidal variables.

We propose a new method of separating OSARGs from non-variable stars and
SRVs. The method employs the position in the reddening-free {\it PL}
diagrams and the characteristic period ratios of these multiperiodic
variables. The {\it PL} relations for the LMC OSARG are compared with the
calculated relations for RGB models along isochrones of relevant ages and
metallicities. We also compare measured periods and amplitudes of the
OSARGs with predictions based on the relations valid for less luminous
solar-like pulsators.
  
Miras and SRVs seem to follow {\it PL} relation of the same slopes in the
LMC and SMC, while for LSP and ellipsoidal variables slopes in both
galaxies are different. The {\it PL} sequences defined by LSP variables and
binary systems overlap in the whole range of analyzed wavebands. We put
forward new arguments for the binary star scenario as an explanation of the
LSP variability and elaborate on it further. The measured pulsation to
orbital period ratio implies nearly constant ratio of the star radius to
orbital distance, $R/A\approx0.4$, as we find. Combined effect of tidal
friction and mass loss enhanced by the low-mass companion may explain why
such a value is preferred.}{Stars: AGB and post-AGB -- Stars: late-type --
Stars: oscillations -- Magellanic Clouds}

\Section{Introduction}
The first attempt to determine period--luminosity (PL) relation for long
period variables (LPVs) was made by Gerasimovi\v{c} (1928), who noticed
that Mira stars with longer periods are on average fainter at visual
wavelengths. This result has been confirmed by subsequent studies (\eg
Wilson and Merrill 1942, Osvalds and Risley 1961, Clayton and Feast 1969),
however the scatter of this period--luminosity dependence turned out to be
very large.

First tight {\it PL} relation for LPVs was discovered for Mira stars at
near infrared (NIR) wavebands (Glass and Lloyd Evans 1981). This {\it PL}
law, based on only 11 Miras in the LMC, was refined by extensive studies of
Feast \etal (1989) and Hughes and Wood (1990). The second, parallel {\it
PL} sequence, occupied by semiregular variables (SRVs), was identified by
Wood and Sebo (1996). This sequence was shifted relative to the Miras'
ridge toward shorter periods by a factor of two.

However, the subject of {\it PL} distribution of LPVs has progressed
rapidly over recent years when large microlensing surveys (MACHO, OGLE,
EROS, MOA) published long-term photometry of huge number of stars. Complex
structure of the {\it PL} distribution was demonstrated for the first time
by Cook \etal (1997), who published {\it PL} diagram for variable stars
detected during the MACHO survey in the LMC. A series of three or four {\it
PL} sequences defined by LPVs can be distinguished in that diagram.

Sharper picture was presented by Wood \etal (1999), who distinguished and
described five {\it PL} sequences (denoted as A--E) in the
period--Wesenheit index plane. Wood (2000) showed similar distribution in
the $\log P$--$K$ diagram. These results were then confirmed by many
studies based on observations originated in various sources (Cioni \etal
2001, 2003, Noda \etal 2002, Lebzelter \etal 2002, Ita \etal 2004,
Groenewegen 2004, Fraser \etal 2005).

Kiss and Bedding (2003, 2004) used OGLE data to reveal new features in the
{\it PL} distribution. They noticed that sequence B consists of two closely
spaced parallel ridges (Ita \etal 2004 denoted the additional sequence as
C$'$). Below the tip of the red giant branch (TRGB) Kiss and Bedding (2003)
found three sequences shifted in $\log P$ relative to stars brighter than
TRGB. It was the definitive proof that stars in the first ascent Red Giant
Branch (RGB) pulsate similarly to objects being in the Asymptotic
Giant Branch (AGB) phase.

The Optical Gravitational Lensing Experiment (OGLE) collected unprecedented
amount of photometric data of stars in the Large and Small Magellanic
Clouds. Both galaxies have been constantly monitored since 1997 and at
present time this is the best and longest available photometric dataset for
analyzing huge number of variable red giants. Our studies on LPVs resulted
in many discoveries, regarding also the {\it PL} relations.

Soszy{\'n}ski \etal (2004a) showed that OGLE Small Amplitude Red Giants
(OS\-ARGs) constitute separate class of variable stars, with different
structure in the {\it PL} plane than ``classical'' SRVs and Miras. We
indicated two previously overlooked {\it PL} relations -- the longest
(${\rm a}_1$) and the shortest (${\rm a}_4$) period sequences followed by
AGB OSARGs. We also suggested a method of empirical division between RGB
and AGB OSARGs fainter than TRGB.

Red giants revealing ellipsoidal modulation caused by binarity were
analyzed by Soszy{\'n}ski \etal (2004b). It was shown that, if true orbital
periods are considered, the {\it PL} relation of ellipsoidal variables
(sequence~E) is a direct continuation of sequence~D occupied by mysterious
Long Secondary Period (LSP) variables. This is a hint that the LSP
phenomenon may be related to binarity, but taking into account available
radial velocity measurements, the secondary component usually must be a low
mass object, possibly former planet. This idea was supported by
Soszy{\'n}ski (2007) who discovered in some LSP variables ellipsoidal-like
and eclipsing-like modulations with periods equal to LSPs.

In Soszy{\'n}ski \etal (2005) we again increased complexity of the PL
distribution of LPVs. Each of the NIR {\it PL} sequences C$'$, C and D in
the LMC (occupied by SRV, Miras and LSP variables) split into two separate
ridges in the period -- optical Wesenheit index plane, what corresponds to
the spectral division into oxygen-rich (O-rich) and carbon-rich (C-rich)
AGB stars. Thus, we found a new photometric method of distinguishing between
these two populations.

In this paper we describe in details the {\it PL} relations of variable red
giants in both Magellanic Clouds. We show new details in the {\it PL} plane
and compare {\it PL} distribution in the LMC and SMC. The paper is
organized as follows. Section~2 gives details of the observations and data
reduction. In Section~3 the {\it PL} relations are presented with a
description of their derivation. A discussion about four types of red giant
variability -- OSARGs, Miras/SRVs, LSPs and ellipsoidal modulation -- is
given in Sections~4--7. Section~8 summarizes and concludes the paper.

\Section{Observations and Data Reduction}
Our analysis is based on the photometry of stars in the LMC and SMC
observed during the second and third phases of the OGLE survey (OGLE-II and
OGLE-III). Both galaxies have been monitored since 1997 with the 1.3-m
Warsaw telescope at the Las Campanas Observatory, Chile, which is operated
by the Carnegie Institution of Washington. During the OGLE-II survey
(1997--2000) the telescope was equipped with the ``first generation''
camera with the SITe ${2048\times2048}$ CCD detector. Details of the
instrumentation setup can be found in Udalski \etal (1997). In 2001 the
telescope equipment was upgraded to a~wide field ${8192\times8192}$ mosaic
camera consisting of eight SITe CCD detectors (Udalski 2003).

In this work we used OGLE-II observations supplemented by the OGLE-III
data. The OGLE-II fields cover about 4.5 square degrees in the central
regions of the LMC and 2.4 square degrees in the SMC. The last observations
used in this analysis were collected in February 2007, so our photometry
spans 10 years. The data have been obtained in two standard {\it I} and
{\it V} filters, but majority of the observations (typically 700--900
points per star) were carried out in the former bandpass. Saturation
occurred for stars brighter than $I=12.5$~mag. In the {\it V}-band we
collected about 50 points per star.

The data were reduced using the standard OGLE pipeline, as described by
Udalski (2003). For details about the transformation of the instrumental
photometry to the standard system and the determination of the equatorial
coordinates of stars the reader is referred to Udalski \etal (2000). The
well calibrated OGLE-II photometry and OGLE-III observations were tied
using magnitudes of several dozen constant stars in the closest
neighborhood around each object.

The $K_S$ and $J$ NIR photometry used in this analysis was taken from 2MASS
All-Sky Catalog of Point Sources (Cutri \etal 2003). We performed
cross-identifica\-tion between OGLE and 2MASS sources with 1~arcsec search
radius, but earlier small shifts to RA and DEC had been added to compensate
for systematic differences between both coordinate systems. We found
confident 2MASS counterparts for 98\% of red stars observed by OGLE. For
further analysis, we included only stars with all four -- $V$, $I$, $J$ and
$K_S$ -- bands available.
\vspace*{9pt}
\Section{Period--Luminosity Relations for Variable Red Giants}
\vspace*{4pt}
A period analysis for all stars with $(V-I)>0.5$~mag and $I<17$~mag (in the
SMC with $I<17.5$~mag) was performed with program {\sc Fnpeaks} developed
by Z.~Ko{\l}aczkowski (private communication). The frequency range searched
was from 0.0 to 0.9 cycles per day. For each light curve we determined the
highest peak in the Fourier spectrum, then the third order Fourier series
was fitted to the observations and subtracted from it. The procedure was
repeated on the residual data until fifteen periods per star were derived.
Moreover, we recorded the amplitudes and signal-to-noise (S/N) parameters
associated with each period.

We constructed {\it PL} diagrams in three variants for both galaxies: $\log
P$--$K_S$, $\log P$--$W_{JK}$, and $\log P$--$W_I$. $K_S$ is the 2MASS
single epoch magnitude with no compensation applied for interstellar
reddening. The single epoch random phase observations cause additional
spread in the {\it PL} relations for large amplitude variables. This effect
has been avoided for Miras and SRVs thanks to converting single epoch
observations into mean magnitudes using complete {\it I}-band light
curves. Details of this method were described by Soszy{\'n}ski \etal
(2005).

$W_{JK}$ and $W_I$ are NIR and optical Wesenheit indices (Madore 1982), \ie
red\-dening free quantities being a linear combinations of selected
magnitudes and colors:
\begin{eqnarray}
W_{JK}&=&K_S-0.686(J-K_S)\\
\noalign{\vskip6pt}
W_I&=&I-1.55(V-I)
\end{eqnarray}

\newpage
\renewcommand{\TableFont}{\scriptsize}
\renewcommand{\arraystretch}{1.05}
\MakeTable{l@{\hspace{4pt}}c@{\hspace{4pt}}l@{\hspace{8pt}}r@{\hspace{4pt}}r@{\hspace{8pt}}r@{\hspace{4pt}}r@{\hspace{8pt}}r@{\hspace{4pt}}r@{\hspace{4pt}}r@{\hspace{8pt}}}{12.5cm}
{Period--Luminosity Relations of LPVs in the LMC}
{\hline
\noalign{\vskip3pt}
 &&& \multicolumn{2}{c}{$K_S=\alpha(\log{P}-2.0)+\beta$} & \multicolumn{2}{c}{$W_{JK}=\alpha(\log{P}-2.0)+\beta$} & \multicolumn{3}{c}{$W_I=\alpha\log^2{P}+\beta\log{P}+\gamma$} \\
\multicolumn{3}{c}{\raisebox{1.5ex}{Sequence}} & \multicolumn{1}{c}{$\alpha$} & \multicolumn{1}{c}{$\beta$} & \multicolumn{1}{c}{$\alpha$} & \multicolumn{1}{c}{$\beta$} & \multicolumn{1}{c}{$\alpha$} & \multicolumn{1}{c}{$\beta$} & \multicolumn{1}{c}{$\gamma$} \\
\noalign{\vskip3pt}
\hline
\noalign{\vskip3pt}
       &     & ${\rm b}_1$ & $-3.34\pm0.04$ & $11.51\pm0.02$ & $-3.81\pm0.05$ & $10.64\pm0.02$ & $-2.648$ & $3.780$ & $13.098$ \\
       & RGB & ${\rm b}_2$ & $-3.58\pm0.04$ & $10.94\pm0.02$ & $-3.96\pm0.04$ & $10.03\pm0.03$ & $-3.396$ & $5.118$ & $12.182$ \\
       &     & ${\rm b}_3$ & $-3.72\pm0.05$ & $10.22\pm0.03$ & $-4.02\pm0.05$ &  $9.32\pm0.02$ & $-3.564$ & $4.473$ & $12.592$ \\
OSARGs &     & ${\rm a}_1$ & $-3.69\pm0.04$ & $11.13\pm0.02$ & $-3.92\pm0.04$ & $10.31\pm0.02$ & $-1.604$ & $0.254$ & $15.782$ \\
       &     & ${\rm a}_2$ & $-3.70\pm0.03$ & $10.59\pm0.02$ & $-3.99\pm0.03$ &  $9.70\pm0.02$ & $-1.669$ & $-0.106$ & $15.731$ \\[-1ex]
       &\raisebox{1.5ex}{AGB} & ${\rm a}_3$ & $-3.80\pm0.04$ & $9.98\pm0.03$ & $-4.04\pm0.04$ & $9.08\pm0.03$ & $-1.556$ & $-0.905$ & 15.878 \\
       &     & ${\rm a}_4$ & $-4.01\pm0.04$ & $9.40\pm0.04$ & $-4.18\pm0.04$ & $8.51\pm0.04$ & $-1.382$ & $-1.839$ & 16.235 \\
\noalign{\vskip2pt}
\hline
\noalign{\vskip3pt}
         &  & C$_{\rm O}$ & $-4.17\pm0.08$ & $12.59\pm0.04$ & $-4.34\pm0.09$ & $11.84\pm0.04$ & $-9.803$ & $33.672$ & $-16.149$ \\[-1ex]
Miras    & \raisebox{1.5ex}{O-rich} & C$'_{\rm O}$ & $-4.35\pm0.07$ & $11.25\pm0.02$ & $-4.67\pm0.07$ & $10.39\pm0.02$ & $-9.169$ & $25.589$ & $-5.158$ \\
and SRVs &  & C$_{\rm C}$ & $-4.07\pm0.11$ & $12.71\pm0.06$ & $-5.19\pm0.11$ & $12.01\pm0.06$ & $-6.618$ & $25.468$ & $-12.522$ \\[-1ex]
         & \raisebox{1.5ex}{C-rich} & C$'_{\rm C}$ & $-4.06\pm0.08$ & $11.39\pm0.04$ & $-4.89\pm0.08$ & $10.39\pm0.04$ & $-3.133$ & $7.278$ & $8.830$ \\
\noalign{\vskip3pt}
\hline
\noalign{\vskip3pt}
 & O-rich & D$_{\rm O}$ & $-4.41\pm0.07$ & $15.05\pm0.05$ & $-4.64\pm0.08$ & $14.43\pm0.05$ & $-5.882$ & $23.707$ & $-10.128$ \\[-1ex]
\raisebox{1.5ex}{LSPs} & C-rich & D$_{\rm C}$ & $-4.38\pm0.25$ & $15.26\pm0.23$ & $-5.09\pm0.28$ & $14.92\pm0.26$ & $0.0$ & $-7.71$ & $33.44$ \\
\noalign{\vskip3pt}
\hline
\noalign{\vskip3pt}
\multicolumn{2}{l}{Ell.} & E & $-3.41\pm0.10$ & $14.48\pm0.06$ & $-3.79\pm0.11$ & $13.92\pm0.06$ & $0.0$ & $-3.94$ & $22.20$ \\
\noalign{\vskip3pt}
\hline
}
\MakeTable{l@{\hspace{4pt}}c@{\hspace{4pt}}l@{\hspace{8pt}}r@{\hspace{4pt}}r@{\hspace{8pt}}r@{\hspace{4pt}}r@{\hspace{8pt}}r@{\hspace{4pt}}r@{\hspace{4pt}}r@{\hspace{8pt}}}{12.5cm}
{Period--Luminosity Relations of LPVs in the SMC}
{\hline
\noalign{\vskip3pt}
 &&& \multicolumn{2}{c}{$K_S=\alpha(\log{P}-2.0)+\beta$} & \multicolumn{2}{c}{$W_{JK}=\alpha(\log{P}-2.0)+\beta$} & \multicolumn{3}{c}{$W_I=\alpha\log^2{P}+\beta\log{P}+\gamma$} \\
\multicolumn{3}{c}{\raisebox{1.5ex}{Sequence}} & \multicolumn1{c}{$\alpha$} & \multicolumn1{c}{$\beta$} & \multicolumn1{c}{$\alpha$} & \multicolumn1{c}{$\beta$} & \multicolumn1{c}{$\alpha$} & \multicolumn1{c}{$\beta$} & \multicolumn1{c}{$\gamma$} \\
\noalign{\vskip3pt}
\hline
\noalign{\vskip3pt}
       &     & ${\rm b}_1$ & $-3.56\pm0.07$ & $11.83\pm0.04$ & $-3.84\pm0.08$ & $11.05\pm0.04$ & $-1.448$ & $0.655$ & $15.482$ \\
       & RGB & ${\rm b}_2$ & $-3.88\pm0.07$ & $11.21\pm0.04$ & $-4.25\pm0.08$ & $10.33\pm0.04$ & $-2.844$ & $4.134$ & $12.874$ \\
       &     & ${\rm b}_3$ & $-4.40\pm0.07$ & $10.14\pm0.05$ & $-4.48\pm0.09$ &  $9.39\pm0.05$ & $-3.583$ & $4.782$ & $12.639$ \\
OSARGs &     & ${\rm a}_1$ & $-3.66\pm0.04$ & $11.55\pm0.02$ & $-4.02\pm0.05$ & $10.74\pm0.03$ & $-0.520$ & $-2.574$ & $18.041$ \\
       &     & ${\rm a}_2$ & $-3.89\pm0.04$ & $10.85\pm0.02$ & $-4.11\pm0.04$ & $10.06\pm0.03$ & $-1.111$ & $-0.912$ & $16.248$ \\[-1ex]
       & \raisebox{1.5ex}{AGB} & ${\rm a}_3$ & $-4.04\pm0.05$ & $10.19\pm0.04$ & $-4.40\pm0.05$ & $9.25\pm0.04$ & $-1.179$ & $-1.090$ & 16.001 \\
       &     & ${\rm a}_4$ & $-4.29\pm0.04$ & $9.52\pm0.03$ & $-4.49\pm0.04$ & $8.89\pm0.04$ & $-0.688$ & $-2.886$ & 16.997 \\
\noalign{\vskip2pt}
\hline
\noalign{\vskip3pt}
         &  & C$_{\rm O}$ & $-4.14\pm0.17$ & $12.90\pm0.05$ & $-4.39\pm0.19$ & $12.14\pm0.05$ & $-6.900$ & $23.341$ & $-6.495$ \\[-1ex]
Miras    & \raisebox{1.5ex}{O-rich} & C$'_{\rm O}$ & $-4.47\pm0.15$ & $11.62\pm0.03$ & $-4.94\pm0.15$ & $10.81\pm0.03$ & $-3.816$ & $8.379$ & $9.106$ \\
and SRVs &  & C$_{\rm C}$ & $-4.22\pm0.16$ & $13.11\pm0.07$ & $-5.26\pm0.14$ & $12.43\pm0.06$ & $-3.991$ & $12.486$ & $3.605$ \\[-1ex]
         & \raisebox{1.5ex}{C-rich} & C$'_{\rm C}$ & $-4.40\pm0.15$ & $11.88\pm0.04$ & $-5.05\pm0.13$ & $10.90\pm0.04$ & $-3.482$ & $7.933$ & $9.362$ \\
\noalign{\vskip3pt}
\hline
\noalign{\vskip3pt}
 & O-rich & D$_{\rm O}$ & $-4.13\pm0.11$ & $15.16\pm0.08$ & $-4.40\pm0.11$ & $14.61\pm0.08$ & $-4.752$ & $19.487$ & $-6.363$ \\[-1ex]
\raisebox{1.5ex}{LSPs} & C-rich & D$_{\rm C}$ & $-4.11\pm0.30$ & $15.39\pm0.30$ & $-4.46\pm0.33$ & $14.77\pm0.30$ & $0.0$ & $-7.00$ & $31.40$ \\
\noalign{\vskip3pt}
\hline
\noalign{\vskip3pt}
\multicolumn{2}{l}{Ell.} & E & $-3.15\pm0.12$ & $14.70\pm0.07$ & $-3.32\pm0.16$ & $14.10\pm0.07$ & $0.0$ & $-3.90$ & $22.43$ \\
\noalign{\vskip3pt}
\hline
}

\begin{figure}[p]
\hglue-11mm{\includegraphics[width=14.5cm]{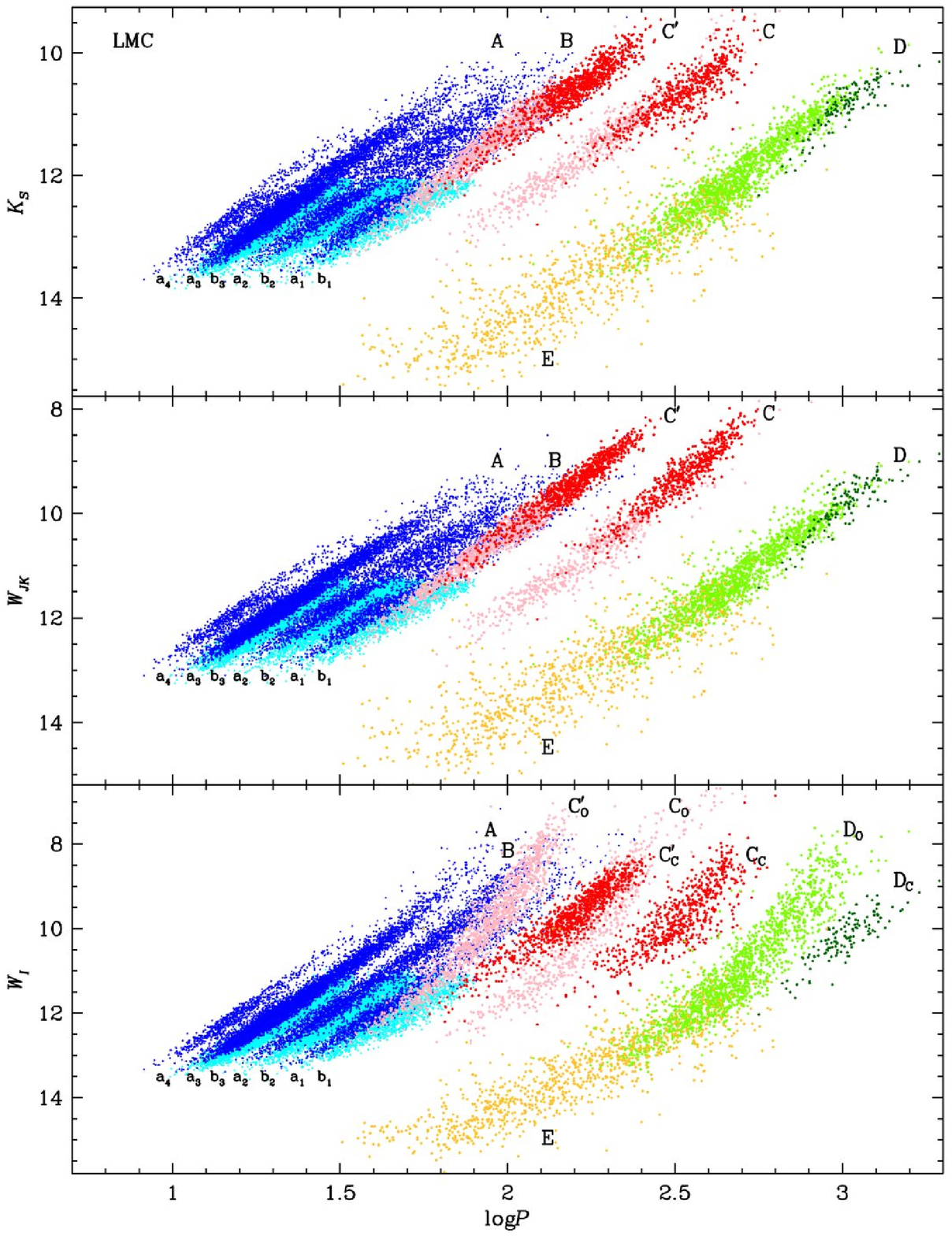}}
\vspace{-1.1cm}
\FigCap{Period--luminosity diagrams of variable red giants in the
LMC. OSARG variables are shown as blue points (RGB as light blue, AGB as
dark blue). Miras and SRVs are marked with pink (O-rich) and red (C-rich)
points. Light and dark green points refer to O-rich and C-rich LSP
variables, respectively. Yellow points indicate ellipsoidal red giants.}
\end{figure}
\begin{figure}[p]
\hglue-11mm{\includegraphics[width=14.5cm]{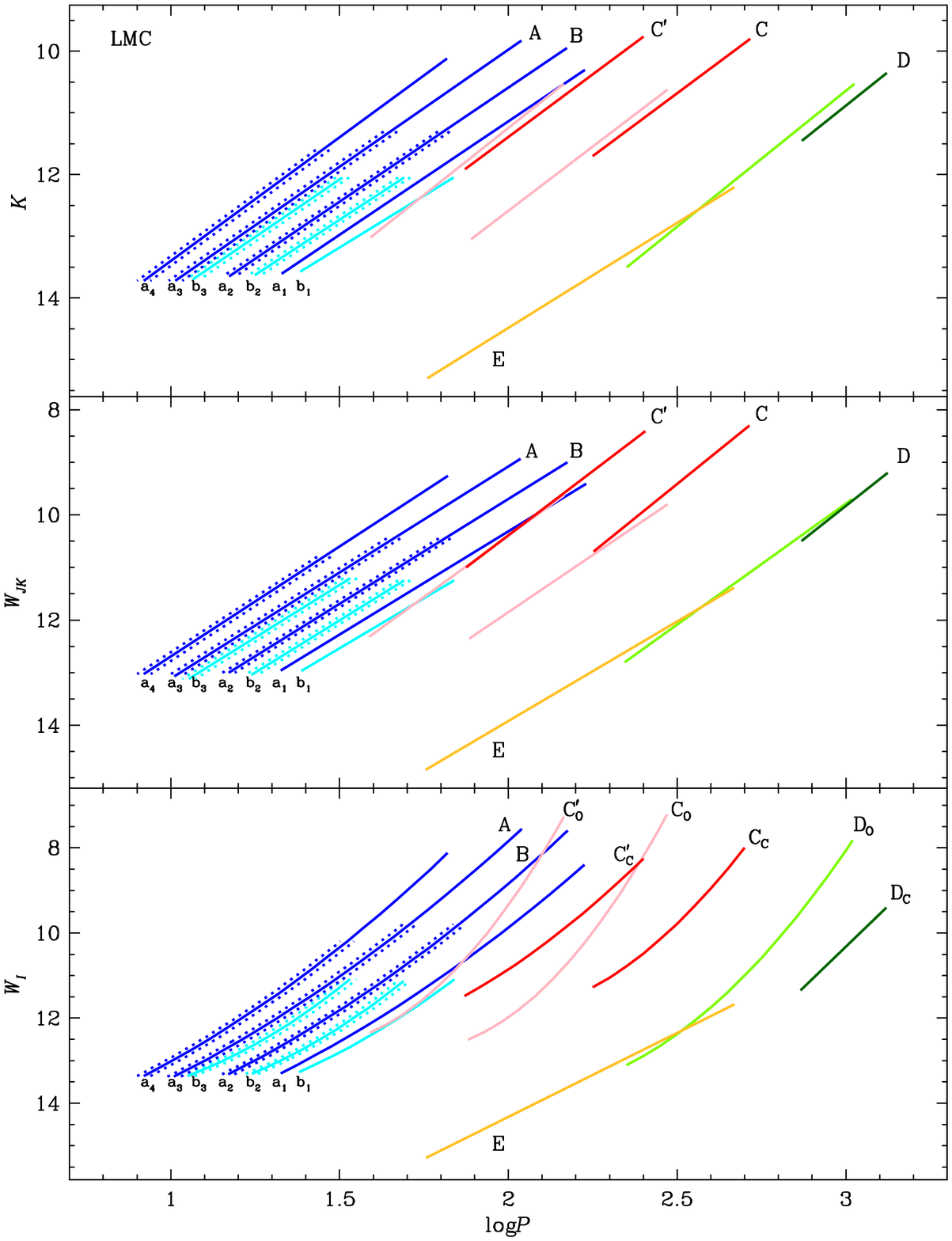}}
\vspace{-0.8cm}
\FigCap{Period--luminosity relations for red giants in the LMC fitted to the sequences shown in
Fig.~1. The colors of lines indicate the same types of stars as in Fig.~1.}
\end{figure}
\begin{figure}[p]
\hglue-11mm{\includegraphics[width=14.5cm]{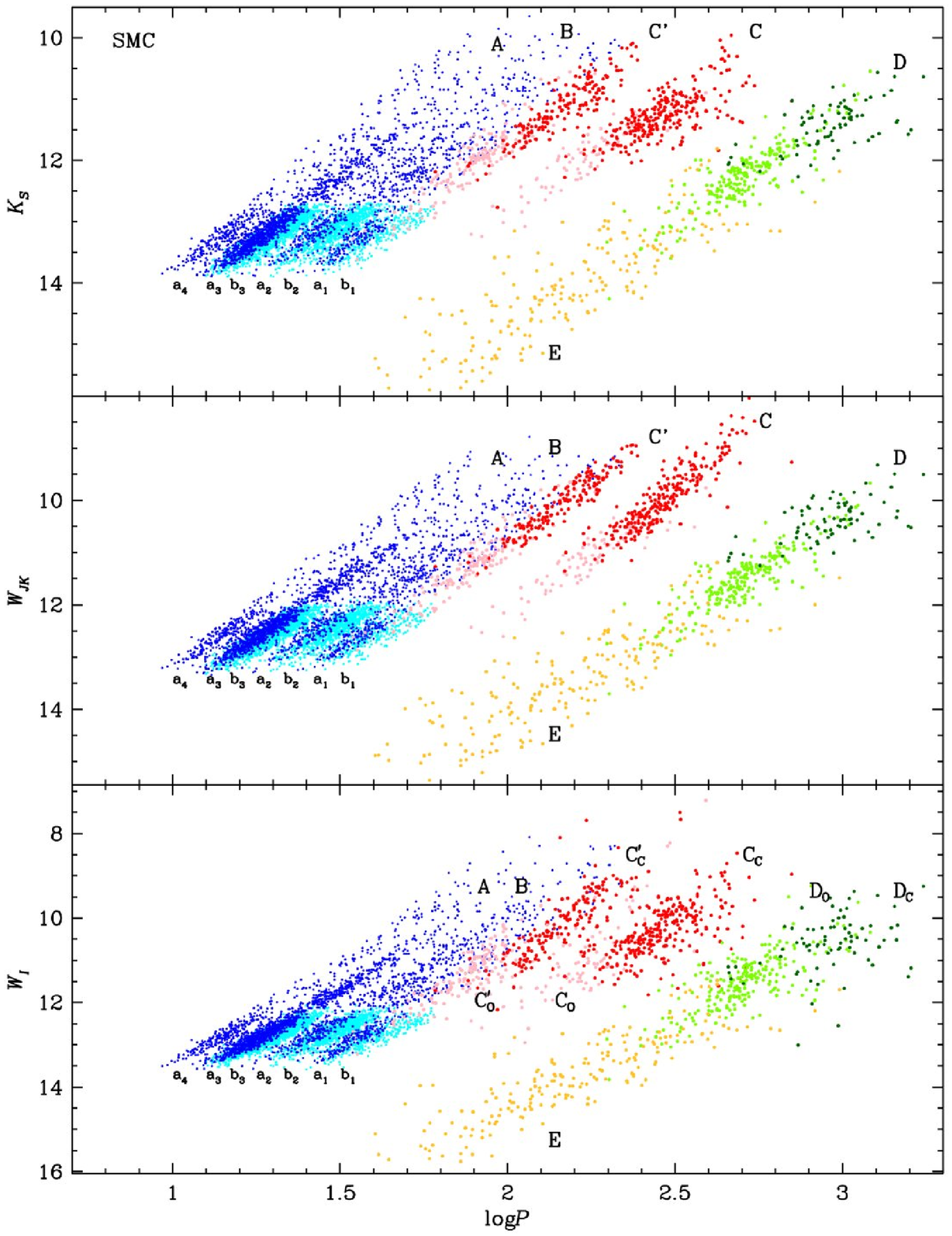}}
\vspace{-0.8cm}
\FigCap{Period--luminosity diagrams of variable red giants in the
SMC. The colors represent the same types of stars as in Fig.~1.}
\end{figure}
\begin{figure}[p]
\hglue-11mm{\includegraphics[width=14.5cm]{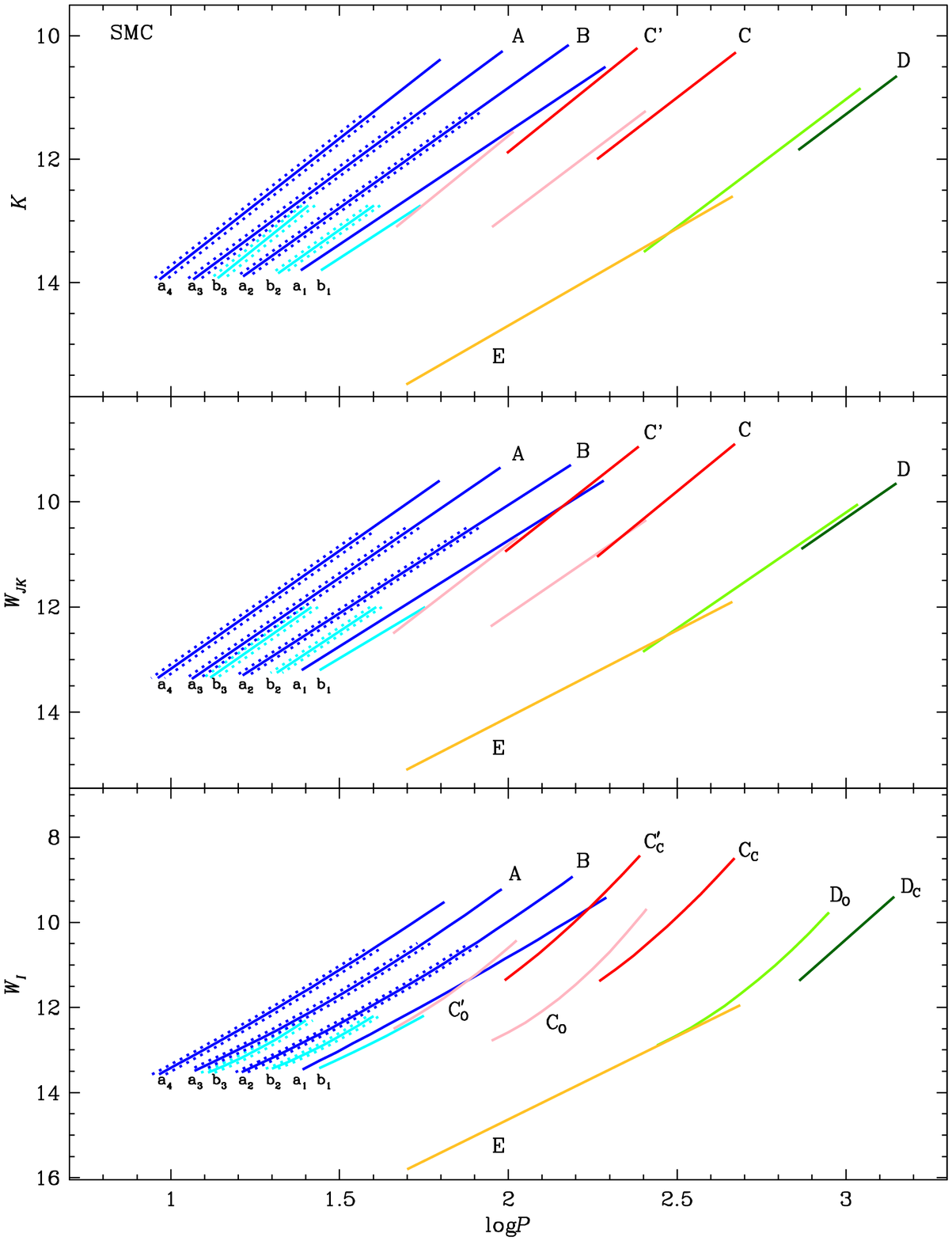}}
\vspace{-0.8cm}
\FigCap{Period--luminosity relations for red giants in the SMC fitted to the sequences shown in
Fig.~3. The colors of lines indicate the same types of stars as in
Fig.~1.}
\end{figure}

The {\it PL} diagrams for LMC and SMC are shown in Fig.~1 and Fig.~3,
respectively. We present here only the stars that have been used to
determine the {\it PL} relations. For Miras and SRVs, LSP and ellipsoidal
variables only one point per star is shown, while for OSARGs more periods
per star may by displayed. Detailed description of the criteria used for
the selection of variables are given in the next four Sections. The LMC PL
sequences appear better defined which may be explained by much larger
sample of objects and by much smaller depth in the line of sight of the LMC
than the SMC.

To derive possibly most precise regression lines we first converted the
{\it PL} diagrams into density maps and found points along the given {\it
PL} sequence where the density of points is the largest. Then, we performed
linear or quadratic least square fit to these points. Note that the
described method produces regression functions in general different than
ordinary least square fits to the {\it PL} points, \ie we usually obtained
steeper lines.  Figs.~2 and~4 shows the functions fitted to the {\it PL}
sequences. The same functions are given in Tables~1 and~2.

At $K_S$ and $W_{JK}$ the {\it PL} relations have been approximated by
linear functions, while $\log P$--$W_I$ laws have usually been fitted with
quadratic functions, because of distinct curvature of these relations. This
non-linearity is associated curvature of the distribution visible in the
color--magnitude $(V-I)$--$I$ diagram. Optical and NIR color--magnitude
diagrams for stars included in Figs.~1 and~3 are shown in Fig.~5. One can
notice different features of $(V-I)$--$I$ and $(J-K)$--$K$ distributions in
both Clouds. Note, that the coefficients of the parabolas fitted to the
period--$W_I$ relations (Tables~1 and~2) have considerable
errors. Nevertheless, one may find these cofficients useful.

\begin{figure}[t!]
\centerline{\includegraphics[width=14cm]{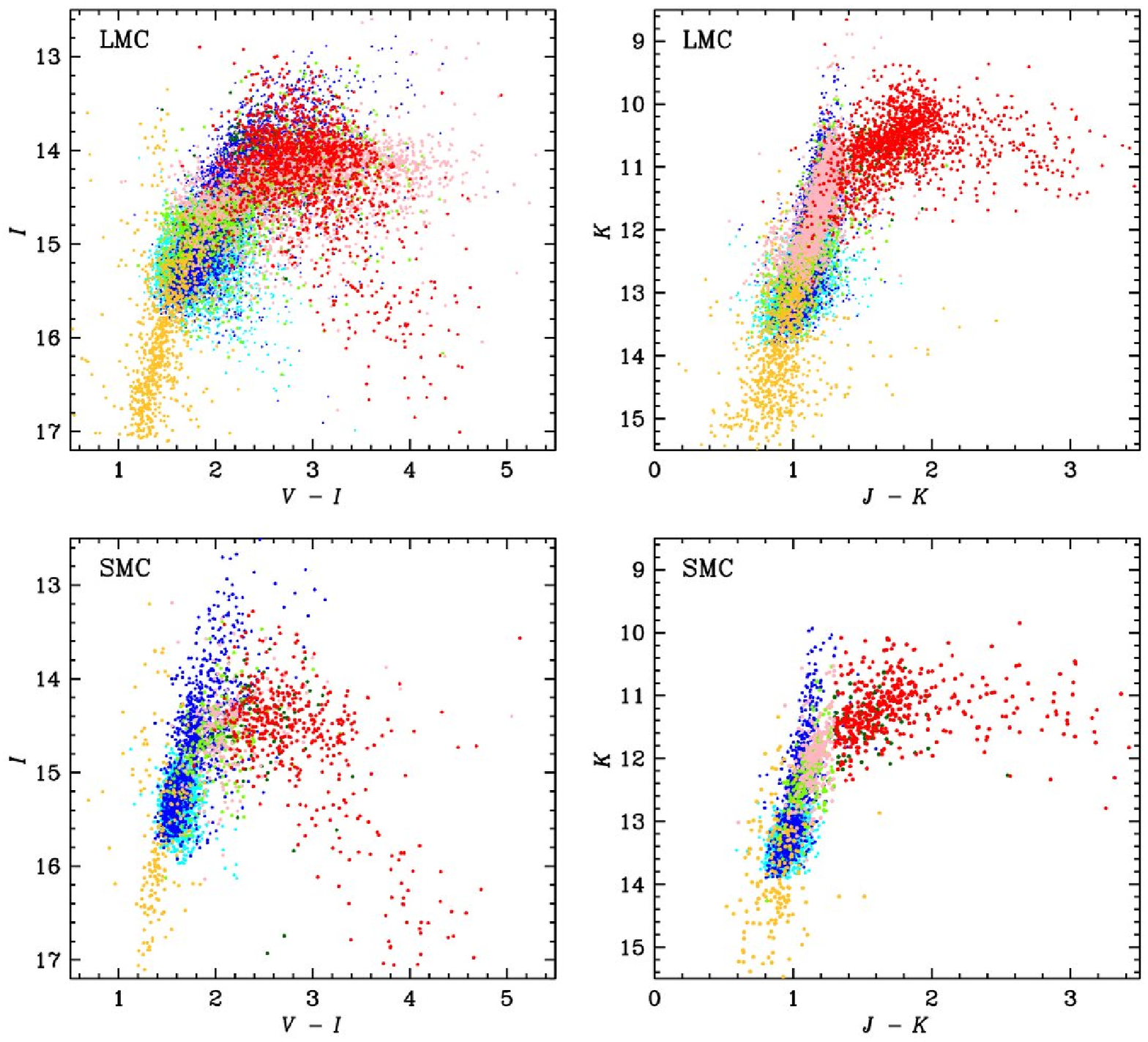}}
\vspace{-0.5cm}
\FigCap{Color--magnitude diagrams for LMC ({\it upper panels}) and SMC ({\it lower
panels}) variable red giants. The colors represent the same types of stars
as in Fig.~1.}
\end{figure}

In the next Sections we describe in more details four classes of variable red
giants visible in the {\it PL} diagrams.

\Section{OGLE Small Amplitude Red Giants}
The name ``OGLE Small Amplitude Red Giants'' (OSARGs) has been proposed by
Wray \etal (2004), who analyzed about 18\,000 of such objects in the
Galactic bulge using OGLE-II photometry. The term ``OSARG'' is not fully
synonymous with ``Small Amplitude Red Variable'' (SARV) from the
classification introduced by Eggen (1977, and references within), \ie stars
with visual amplitudes smaller than 0.5~mag. As shown by Soszy{\'n}ski
\etal (2004a) the OSARG variables cannot be distinguished from
``classical'' SRVs (hereafter just SRVs) using solely amplitudes of
variations, because this is an overlapping property of both
groups. Moreover, Eggen's SARVs include wide variety of stellar types,
while OSARG variables constitute a separate class of variable stars
(Soszy{\'n}ski \etal 2004a).

It is known that all red giants of type K5 and cooler are variable in
brightness, and amplitude of variations increases with decreasing
temperature of the stars (Edmonds and Gilliland 1996, Henry \etal 2000).
Using OGLE photometry we can detect stars with {\it I}-band peak-to-peak
amplitudes as small as 4~mmag.

\Subsection{Selection of OSARGs}
Since OSARGs with the smallest detectable amplitudes are mixed with
constant stars (strictly speaking, with stars having amplitudes too small
to be visible in our data), it is not an easy task to separate both groups.
Likewise, the longest period sequences of OSARGs overlap with {\it PL}
ridge of SRVs, thus a separation of the two populations is in general
problematic. Any of simple diagrams that can be prepared using available
parameters of stars (periods, magnitudes, colors, amplitudes) did not
enable us to precisely separate long-period OSARGs from SRVs, and faintest
OSARGs from non-variable stars.

To solve this problem we worked out a new method of OSARG variables
detection. Our algorithm takes into account a star position on the PL
diagrams and characteristic period ratios of these multiperiodic variables.
Our basic tool is the diagram $\Delta\log P_L$--$P_S/P_L$, where $P_S$ and
$P_L$ is any pair of periods (shorter and longer, respectively) selected
from the most significant periodicities of the given star, and $\Delta\log
P_L$ is a horizontal distance of $\log P_L$ from selected sequence in the
selected {\it PL} plane (in our analysis from sequence~A).

\begin{figure}[htb]
\centerline{\includegraphics[width=12.2cm]{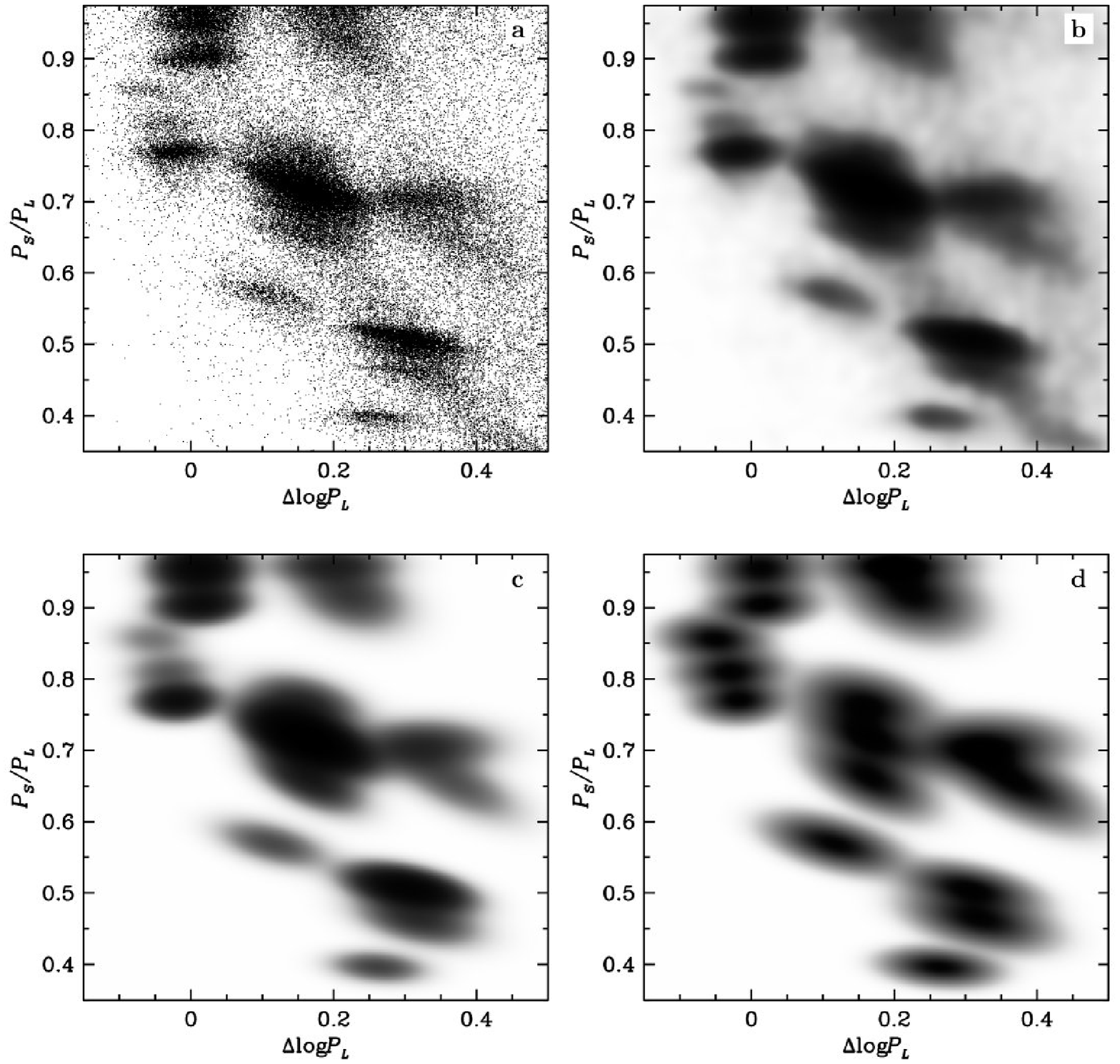}}
\FigCap{Diagram $\Delta\log P_L$--$P_S/P_L$ for OSARG variables in the
LMC. $P_S$ and $P_L$ are, respectively, any shorter and longer periods of
a given star. $\Delta\log{P_L}$ is a horizontal distance from sequence~A
in $\log P$--$W_I$ plane. See text for detailed description.}
\end{figure}
Fig.~6a shows such a diagram for OSARGs tentatively selected from our data.
$\Delta\log P_L$ in this diagram were measured in the $\log P_L$--$W_I$
plane, but very similar picture was independently obtained using $\log
P_L$--$W_{JK}$ diagram. Distinct groups of points are fingerprints of the
OSARG variability. In order to utilize this distribution for selecting
OSARGs, we smoothed it with a Gaussian filter and derived density map
visualized in Fig.~6b.

Then, we fitted two-dimensional Gauss function to every group of points
visible in Fig.~6b, and prepared artificial density map which is shown in
Fig.~6c. In this way we rejected points which occurred in the $\Delta\log
P_L$--$P_S/P_L$ diagrams by chance from spuriously detected periods. The
last step of our procedure was normalization of our map, \ie setting up the
maxima of all fitted Gaussian at the same level (Fig.~6d). Thanks to that
every pair of periods was treated with the same weight, regardless of which
{\it PL} relations corresponded to these periods. Two such maps as shown in
Fig.~6d have been prepared -- one with $\Delta\log P_L$ measured in the
$\log P_L$--$W_I$ plane, and the second in the $\log P_L$--$W_{JK}$
diagram.

The procedure of selecting OSARG variables was performed as follows. For
each star we examined each pair of periods (selected from 15 derived
periods) for which S/N parameter was larger than 3.0. We derived ratios of
both periods and $\Delta\log P_L$ parameters in the $\log P_L$--$W_I$ and
$\log P_L$--$W_{JK}$ planes. Then, for subsequent Gaussians shown in
Fig.~6d we calculated values in respective ($\Delta\log P_L$, $P_S/P_L$)
points, thus only points close to the maximum of the given Gaussian
produced values significantly larger than zero. Additionally, we took the
product of the values obtained with the two Wesenheit indices so that the
values for accidental points were close to zero.

We co-added every such derived value for each Gaussian and each pair of
periods. Resulting sum is an indicator of the OSARG variability. For the
best candidates for OSARG variables this quantity is larger than 10. For
our purposes we set the threshold of the OSARG indicator to be larger than
3. Our method allows effective discrimination of OSARGs from non-variable
stars, as well as from SRVs.

\subsection{Period--Luminosity Relations for OSARGs}
The OSARGs constitute probably the most numerous class of variable stars in
the Magellanic Clouds. We selected about 17\,000 such objects in the LMC
and about 3500 in the SMC, which is more than a number of Cepheids, RR~Lyr
stars, eclipsing binaries and other types of variable red giants. OSARG
variables brighter than TRGB ($K_S=12.05$~mag in the LMC, $K_S=12.7$~mag in
the SMC) were obviously recognized as AGB stars, while below TRGB we
separated RGB and AGB stars using the feature noticed by Soszy{\'n}ski
\etal (2004a). If any of the periods falls on the shortest-period sequence
${\rm a}_4$, the star was recognized as AGB object. Otherwise we marked the
OSARG as RGB star, although small contribution of AGB may still exist among
these group. In Figs.~1--5 RGB OSARGs are marked with light blue points and
lines, while AGB OSARGs are drawn in dark blue.

As shown by Soszy{\'n}ski \etal (2004a) OSARGs follow a series of three
(RGB stars) or four (AGB stars) narrow sequences, spreading over periods
ranging from 8 to 160 days. In this paper we use labels introduced by
Soszy{\'n}ski \etal (2004a): ${\rm a}_1$~--~${\rm a}_4$ for AGB and ${\rm
b}_1$~--~${\rm b}_3$ for RGB OSARGs. The {\it PL} relations of RGB OSARGs
are shifted in $\log P$ relative to AGB OSARGs, what can be explained by
the temperature difference of both populations at the same luminosity (Kiss
and Bedding 2003). Soszy{\'n}ski \etal (2004a) also showed that the third
sequence (\ie sequences~${\rm a}_3$ and ${\rm b}_3$ or Wood's sequence~A)
is split into three ridges, what manifests itself in the period -- period
ratio diagram with distinct groups of points around $P_S/P_L\approx0.9$ and
$P_S/P_L\approx0.95$. We marked these two additional {\it PL} relations
with dotted lines in Figs.~2 and~4. We observe this feature for RGB and AGB
stars located below the TRGB and for some AGB objects brighter than the
TRGB. For the brightest stars this split is not visible in our data.

Our new analysis revealed that the same feature may also concern other {\it
PL} sequences of OSARG variables. Similar clouds for period ratios 0.9 and
0.95 are clearly visible for the second sequence (${\rm a}_2$ and ${\rm
b}_2$). For the fourth sequence (${\rm a}_4$) this property is marginally
visible, what can be an effect of smaller number of stars that follow this
shortest period relation. Nevertheless, we also drew the additional dotted
lines in the {\it PL} diagrams for this ridge. For the first, longest
period sequences (${\rm a}_1$ and ${\rm b}_1$) we cannot detect any
distinct group of points close to the period ratio 0.9 or~0.95.

It seems that {\it PL} sequences for OSARG variables are best separated in
the $\log P-W_I$ diagram. This plane also reveals the largest differences
between red giants in the LMC and SMC. The {\it PL} sequences in the LMC
are significantly steeper and are more curved than those in the SMC. This
difference may be directly related to different distribution of points in
the color--magnitude ($V-I$, $V$) diagrams given in Fig.~5, what may
reflect differences in ages and metallicity of red giants between the SMC
and LMC. In both galaxies, the slopes of the OSARG {\it PL} relations
increase with decreasing periods.

It is natural to associate the {\it PL} sequences with individual modes of
stellar oscillations. Identification of the modes would yield new
constraints on models of red giants as well as on star formation in the
Magellanic Clouds. As an illustration, we considered the ${\rm b}_1$, ${\rm
b}_2$, and ${\rm b}_3$ sequences in the LMC, assuming that they correspond
to first three radial modes.

Salaris and Girardi (2005) showed that the upper part of the RGB in the LMC
is composed of stars with heavy element abundance in the wide range, from
${\rm [Fe/H]}=-1$ to $-0.3$, which corresponds to age between 1 and
11~Gy. However, for most of the objects ${\rm [Fe/H]}=-0.65\pm0.1$ and
there is a gap in star formation between 7~Gy and 9~Gy.

\begin{figure}[htb]
\centerline{\includegraphics[width=12cm]{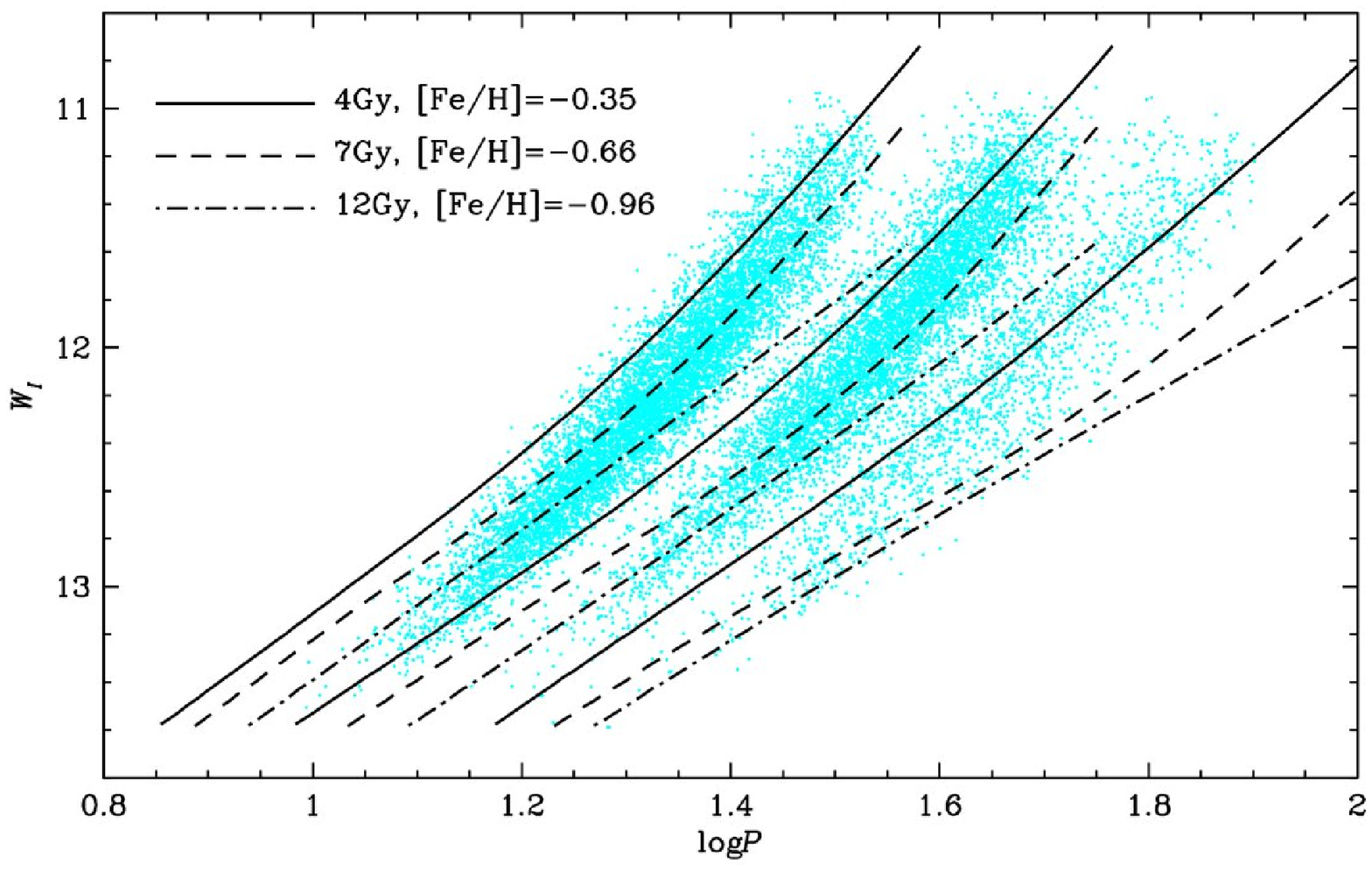}} 
\FigCap{Sequences ${\rm b}_1$, ${\rm b}_2$, and ${\rm b}_3$ of the 
{\it PL} relation for the RGB OSARGs of the LMC (see Fig.~1) compared with
relations calculated on the basis of the selected BaSTI isochrones for the
first three radial modes. Ages and metal abundance parameters are given in
legend. The initial masses in the order of increasing age are equal
1.26~\MS, 1.02~\MS, and 0.85~\MS. Mass loss in the RGB is included in the
model calculations and the corresponding masses at the RGB-tip are 1.13~\MS,
0.94~\MS, and 0.79~\MS.}
\end{figure}
Our theoretical {\it PL} relations were based on the BaSTI (Pietrinferni
\etal 2004) isochrones. For each model, we recalculated envelope structure
and determined the radial mode periods. The $W_I$ index was calculated from
$V$ and $I$ data from BaSTI assuming 18.45~mag for the distance modulus. A
comparison of the selected theoretical and observed sequences is shown in
Fig.~7. Note that fitting requires younger and more metal abundant objects
at higher luminosities. The agreement between observed and calculated
relations, though not perfect seems encouraging. Detailed comparison
requires simulation of the the LMC RGB population. The result will
certainly differ somewhat from simulation based in the CMD data presented
in Fig.~5 of Salaris and Girardi (2005) and reconciling the results will
require new modeling. It is likely that the four ${\rm a}_k$ sequences
correspond to first four radial modes.

\begin{figure}[p]
\centerline{\includegraphics[width=13cm]{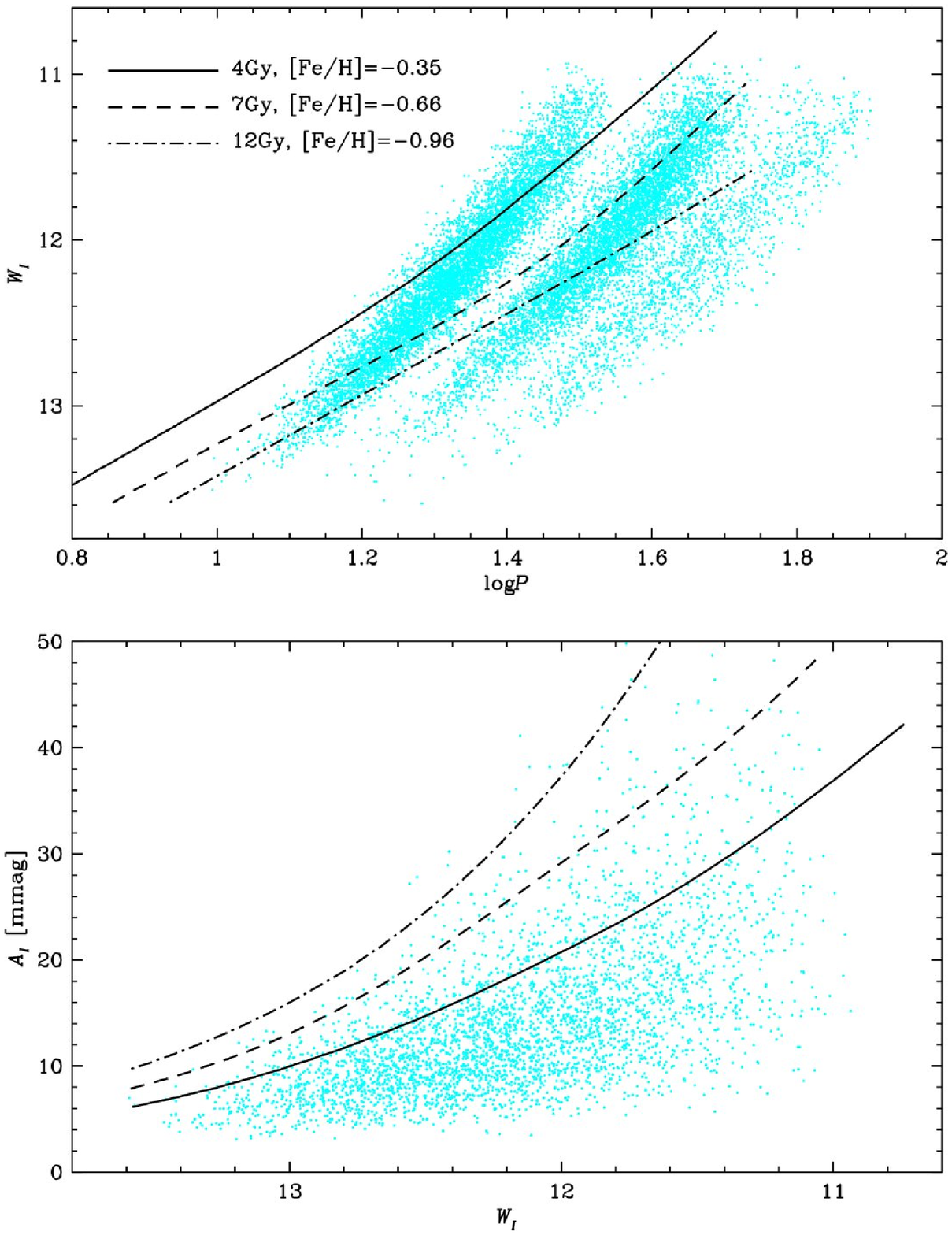}}
\vspace*{-4mm}
\FigCap{The {\it upper panel} shows the same data as in Fig.~7 but the 
lines refer to periods ($P_{\rm max}=\nu_{\rm max}^{-1}$) corresponding to
maximum amplitude of solar-like oscillations calculated with Eq.~(3) for
the same isochrones as in Fig.~7. Dots in the {\it lower panel} shows the
mode amplitudes for the RGB OSARGs of the LMC. The lines were calculated
with Eq.~(4) for the same isochrones as in the {\it upper panel}.}
\end{figure}

\subsection{OSARGs as Solar-Like Pulsators}
It is not yet known how pulsation observed in RGB and AGB stars are
excited. Do they result from an intrinsic mode instability or rather, like
in the sun, from a stochastic energy input to modes which are intrinsically
stable. Both options have been considered Dziembowski \etal (2001) and by
Bedding (2003). In both, there is a crucial role of convection which makes
the problem difficult for theory. Phenomenological arguments, based on
structure of power spectra, in favor for the solar-like option in the case
of the SRVs were first given by Christensen-Dalsgaard \etal (2001) and
supported by Bedding (2003). Power spectra of OSARGs resemble those of
SRVs. Only the amplitudes are smaller. Thus, the solar-like option appears
even more probable for OSARGs.

Stello \etal (2007) reported recently results of a multi-site search for
solar-like oscillations in red giants in the open cluster M67. They did not
report individual mode frequencies but only excess of power in some
individual objects near location predicted by the scaling relation based on
the solar data,
$$\nu_{\rm max}=\frac{(M/\MS)(T_{\rm eff}/{\rm T}_{{\rm eff}\odot})^{3.5}}
{(L/\LS)}\times3050~\mu{\rm Hz}\eqno(3)$$

Bedding and Kjeldsen (2003) showed that this relation approximately
describes location of the highest peaks in oscillation spectra of nearby
dwarfs, subgiants and giants. Stello \etal (2007) also compared the
measured excesses of power with the scaling relation for peak amplitudes
$$A_\lambda=\frac{L/\LS(5.1\pm0.3)~\mu{\rm mag}}{(\lambda/550~{\rm
nm})(T_{\rm eff}/{\rm T}_{{\rm eff}\odot})^2(M/\MS)}\eqno(4)$$
derived by Kjeldsen and Bedding (1995). Only for the objects lying in the
lower RGB ($L/\LS<30$) they were able to find measurable excesses. The
values are crudely consistent with the prediction. For brighter object they
could derive only upper limits.

The RGB OSARGs are considerably brighter than the M67 red giants for which
the excess have been measured. At $W_I=12$~mag their luminosity is between
1.5 and $2\times10^3\LS$. How the two scaling relations work at these high
luminosities is shown in Fig.~8. We used the same isochrones as adopted in
the {\it PL} relations shown in the previous Section, to calculate $P_{\rm
max}=1/\nu_{\rm max}$ using Eq.~(3) and $A_I$ using Eq.~(4). In the latter,
we adopted $\lambda=810$~nm.

In the upper plot we may see that maximum of power corresponds to the first
and second overtone of radial modes. We may conclude that there is a crude
agreement with data. At lower luminosity the calculated lines run well
above the ${\rm b}_1$ sequence. However, we do not take it as an evidence
against the solar-like interpretation of the OSARGs, because in all
solar-like pulsators modes are seen in a rather wide range of
frequencies. On the other hand, since in various stars low-order modes are
excited by the other type of driving, even a very good agreement could not
be regarded as a proof that the OSARGs are indeed solar-like pulsators.

As for the amplitude, the agreement is worse, as the lower panel of Fig.~8
shows. Again, the agreement should not be seen as evidence against the
solar-like nature of the OSARGs. Eq.~(4) is not based on any solid
theory. In fact, a more recent study of stochastic excitation by Samadi
\etal (2005) suggests the $(L/M)^{0.7}$ amplitude scaling, instead of $(L/M)$
adopted in Eq.~(4). Slower amplitude rise is indeed more consistent with
data. At this point, we just want to stress that the OSARGs are likely to
provide the best constraints on models of stochastic excitation over a wide
range of stellar parameters.

\Section{Miras and Semiregular Variables}

Mira stars and SRVs are regarded as two different types of LPVs (Kholopov
\etal 1985), although it seems that both groups represent a continuum. In
the period--NIR luminosity diagrams these objects obey two parallel PL
relations (sequences~C and~C$'$) which are thought to represent fundamental
and first overtone modes of pulsations. Soszy{\'n}ski \etal (2005) studied
3200 Miras and SRVs in the LMC and noticed important feature: in the
period--$W_I$ diagram each {\it PL} sequence splits into two well separated
ridges what corresponds to spectral division into O-rich and C-rich AGB
stars.

In this paper we used the sample of Miras and SRVs in the LMC selected by
Soszy{\'n}ski \etal (2005). We slightly corrected this list by applying the
OSARG indicator described in Section~4.1. This time we left on the list
only variables with very small value of this parameter, smaller than
0.5. In the SMC we selected Miras and SRVs in the same manner as in the LMC
and found in total 630 objects. From our fit we excluded stars redder than
$(J-K_S)=2.3$~mag, \ie objects obscured by dust, and stars populating dim
sequence between C and C$'$ (see discussion below).

In our diagrams Miras and SRVs are indicated with pink (O-rich) and red
(C-rich) points and lines. We plotted only one point per star corresponding
to the most significant period. In the LMC the division into O- and C-rich
variables was done in the period--$W_I$ diagram, where both populations can
be easily distinguished. However, the most striking difference between {\it
PL} relations of LPVs in the LMC and SMC again occurs in the period--$W_I$
plane. In the SMC O-rich and C-rich variables are much worse separated than
in the LMC. Besides, there appears to be almost no bright O-rich Miras and
SRVs in the SMC, what is a well known effect of lower
metallicity. Therefore, in the SMC we used another widely used tool for
distinguishing between O-rich and C-rich red giants -- $(J-K_S)$
colors. Stars bluer than $(J-K_S)=1.3$~mag have been recognized as O-rich
giants, while the remaining objects as C-rich stars.

Let us focus on sequence C$'$ occupied by probable first overtone SRVs. At
the NIR domain this sequence is roughly parallel to OSARG {\it PL}
relations, what was an argument for recognizing OSARGs as the same type of
variables as Miras and SRVs, but pulsating in higher overtones (Wood \etal
1999).  However, the same stars plotted in the $\log P$--$W_I$ diagram
reveal completely different behavior. The {\it PL} law of O-rich SRVs
crosses the OSARG sequences, while C-rich variables are in different
position relative to OSARGs. This discrepancy between NIR and optical {\it
PL} diagrams suggests that OSARGs represent a different, presumably less
evolved, population of stars than SRVs and Miras. To check this, we looked
for OSARGs showing modes belonging to sequence~C, which is defined by
former types of variables.

Among 17\,000 of stars in the LMC classified as OSARGs, we found only a few
objects with distinct periodicity associated with sequence~C. However, for
these objects our classification as OSARGs is uncertain, because none of
the secondary periods fall on sequence~A, which belongs solely to
OSARGs. It is probable that the OSARG signature in these stars is caused
either by spurious periods or periods are real, but in fact associated with
higher overtones of more evolved objects. In conclusion, we did not find
any convincing examples of variables belonging to both classes: SRVs and
OSARGs.

We noted that the secondary periods for some SRVs in both Clouds seem to
follow an additional {\it PL} sequence located between sequences C and
C$'$. Moreover, Soszy{\'n}ski \etal (2005) mentioned about a dim sequence
located in the same place, but defined by the primary periods of some small
amplitude SRVs (these stars are not shown in Figs.~1 and~3). If this
sequence is real and corresponds to a radial mode, then sequence~C cannot
correspond to the fundamental mode. This would imply a revision of the mode
identification in Miras and SRVs. In such a case one can expect that the
fundamental-mode oscillations would be observed for periods by a factor of
about two longer than those associated with sequence~C. Indeed, we have
found a number of SRVs with secondary periods in this region of the PL
diagram, \ie between sequences C and D. Unfortunately, binary systems
(sequence E with halved periods) are also expected to populate this region,
and we would not be able to distinguish between pulsation and ellipsoidal
variations of SRVs in binary systems.

There is also a number of SRVs -- mostly carbon stars -- with secondary
periods shorter than those from sequence C$'$. These periods could be
associated with pulsation modes higher than the first overtone, however
this problem needs further exploration.

We have already noted that the largest differences in the {\it PL}
relations between the two Clouds are seen in the $\log P$--$W_I$
plane. This conclusion concerns especially O-rich stars which populate
different range of magnitudes and the {\it PL} relations differ in
slopes. The sequences in $K_S$ and $W_{JK}$ are more similar. In
particular, the slopes of the {\it PL} relations are the same within the
error limits in both environments. To derive the zero-point differences
between {\it PL} laws, we assumed the same slope for the LMC and the SMC
stars. For the whole sequence~C and for O-rich sequence C$'$, we obtained
similar vertical distances of about 0.35~mag in $K_S$ band and about
0.4~mag in $W_{JK}$. It is in agreement with the previously derived values:
\eg 0.4~mag for dereddened $K_0$ magnitudes obtained by Wood (1995) or
0.36~mag derived by Ita \etal (2004) for sequence~C in $K_S$ (not corrected
for extinction). For carbon stars of sequence C$'$ the zero-point
difference is considerably larger, that is 0.43~mag in $K_S$ and 0.47~mag
in $W_{JK}$. Note that the last value is very close to the estimated
difference between distance moduli of the LMC and SMC (\eg Udalski \etal
1999).

\Section{Long Secondary Periods}
The longest period sequence in our {\it PL} pattern (sequence~D) is
associated with the most mysterious phenomenon connected with red giants
variability.  Long Secondary Periods (LSPs) represent the only unexplained
type of large amplitude stellar variability known today. Wood \etal (2004)
discussed possible explanations of the LSPs and concluded that most
plausible mechanism that may cause LSPs is gravity mode
excitation. However, recently Soszy{\'n}ski (2007) presented observational
arguments for binary origin of the LSPs. In this scenario dust and gas
originated in stellar wind form a cloud orbiting the red giant causing a
periodic obscuration. Radial velocity measurements (Hinkle \etal 2003, Wood
\etal 2004) are consistent with both low-mass companion and non-radial
oscillations. However, our photometric data contain new hints supporting
the former option.

Our samples of LSP variables were initially selected using their position
in the $\log P$--$W_{JK}$ diagrams. After a visual inspection, we chose
about 1600 LMC and 300 SMC distinct LSP stars with characteristic
``eclipsing-like'' light curves. The division into O-rich and C-rich stars
in the LMC was based on the position in the $\log P$--$W_I$ diagram. For
the SMC stars, just like in the case of Miras and SRVs, we relied only on
the $(J-K_S)$ colors.

The strong argument for a binary scenario is that the sequence~D overlaps
with sequence~E, formed by red giants in binary systems, and seems to be a
continuation of this ridge toward brighter stars (Soszy{\'n}ski
\etal 2004b). It is striking that this behavior appears in all analyzed PL
planes: $\log P$--$K_S$ (Derekas \etal 2006), $\log P$--$W_I$, $\log
P$--$W_{JK}$ for both Clouds. Moreover, we found that both sequences seem
to match each other at other bandpasses: $J$, $I$ and even $V$. If the LSP
phenomenon were caused by a stellar oscillation, it would have to be an
unlikely coincidence.

In contrast to sequences~C and~C$'$, the slopes of sequence~D in the NIR
domain are different in both galaxies. The LSP variables in the SMC define
ridge of shallower slope than in the LMC. It is interesting that the same
behavior exhibits sequence~E defined by close binaries. We take this fact
as another argument supporting binary explanation of the LSP phenomenon.

The name ``long secondary periods'' is somewhat misleading, because quite
often amplitude of the LSP variability is higher than that associated with
the short periods. Most of LSP stars are OSARGs both of RGB and AGB
type. However, in the latter case the amplitudes are much higher. Thus our
sample of objects showing LSP consist predominantly of the AGB OSARGs.

OSARG variables with the LSP modulation do not follow exactly the same {\it
PL} relations as non-LSP stars. The larger LSP amplitudes, the larger
discrepancy between {\it PL} sequences. Stars with the largest amplitudes
of the LSP modulations have distinctly smaller mean luminosities compared
to their non-LSP counterparts. Besides, there appears to be a correlation
between the LSP amplitudes and the minimum periods of OSARG oscillations.

An interesting issue concerning LSPs is a relative number of red giants
exhibiting this phenomenon. Typical values found in the literature are
25--30\%. However, our examination shows that this proportion depends on
the lower amplitude of the LSP variability concerned in analysis. Derekas
\etal (2006) indicated that maximum possible amplitudes of the LSP
modulation increase with luminosity of stars, but the smallest amplitudes
are at the detection limit along the whole range of magnitudes. If the
minimum detected amplitudes are around 10~mmag, than the LSPs appear for
about 30\% of LSPs. If the limit for amplitude is decreased to 5~mmag, the
proportion of stars with LSPs reaches 50\%. We can also consider the stars
with LSP variations of amplitudes larger than pulsational (OSARG, SRV)
amplitudes. Then, we detect the LSP modulation in about 30\% of the whole
sample.

Adopting the binary star scenario, we have to explain why the {\it PL}
relation for the LSP is nearly parallel to {\it PL} relations for radial
pulsation modes and why the phenomenon is so common. The near parallel run
of the {\it PL} relations, translates to almost constant ratio of star to
orbit radii, $R/A$. Combining the Kepler law with the expression for radial
mode periods, we obtain
$$\frac{R}{A}=0.24(1+q)^{-1/3}Q_k^{-2/3}\left(\frac{P_k}{P_{\rm
orb}}\right)^{2/3}\eqno(5)$$ 
where $q$, which we will neglect, is the secondary companion to total mass
ratio and $Q_k$ is the pulsation constant. In Section~4.2, we associated
the sequences ${\rm b}_k$ ($k=1,2,3$) of the {\it PL} relations for the RGB
OSARGs with the first three radial modes and suggested the same association
with sequences ${\rm a}_k$ ($k=1,2,3$) for the AGB OSARGs. We follow it
here and, in agreement with the adopted scenario, we set $P_{\rm
orb}=P_{\rm LSP}$. The {\it PL} relation for LSP variables is defined
primarily by the AGB OSARGs, thus we should use periods from the ${\rm
a}_k$ sequences. Unfortunately, we do not have the $Q_k$ values for the AGB
star. However, we noted in our RGB star models that $Q_2$ values change
only little along the isochrones. In the considered models, the values rise
from 0.024 to 0.025. We adopted the upper value and in the Petersen diagram
identified the $P_S/P_L=0.05$ as the $P_3/P_{\rm orb}$ ratio. With these
numbers, Eq.~(5) yields $R/A=0.4$.

What may keep the hypothetical low-mass companion in a narrow range of
fractional distances away of stellar surface? Stars expand through the
whole RGB and most of AGB phase. However, mass loss causes that orbits of
their low-mass companions expand too. At typically adopted mass loss rates,
the $R/A$ ratio increases. As long as its value is well below one, tidal
effects are negligible and changes of the orbital radius are determined
only by mass loss from the star (we ignore here possible interactions with
other companions). The rate of $A$ decrease due to tidal friction is
proportional to $(R/A)^7$ (\eg Zahn 1977), so that once tidal effects
become important, the sign of the temporal derivative of $A$ may quickly
reverse and it is usually assumed (\eg Villaver and Livio 2007) that the
companion is engulfed into the red giant envelope. However, the tidal
interaction affects also the star and it does not seem unreasonable to
assume that it enhances mass loss. If the mass loss rate becomes
sufficiently high the orbital distance my start to increase again and the
companion migration toward the stellar surface may be halted.  Stellar
radius will then continue to increase but the $R/A$ ratio will stay nearly
constant at the value, which we estimated to be about 0.4.

Since about 30--50\% of red giants exhibits the LSP variability and
certainly there must be a range of unfavorable inclinations for its
detection, we must postulate that nearly all red giants have their low mass
companions. This would have far reaching consequences for our views on
frequency planetary systems in the universe. We are aware of problems that
the proposed scenario poses. However, the problems, with an alternative
explanation of the LSP that postulates g-mode excitation seem even harder
to overcome.

\Section{Ellipsoidal and Eclipsing Red Giants}

The last class of LPVs represented in the rich structure of the {\it PL}
distribution are red giants in the binary systems -- ellipsoidal and
eclipsing variables. Wood \etal (1999) suggested that sequence~E is made up
of contact binaries. It was confirmed by Ruci{\'n}ski and Maceroni (2001),
who analyzed long-period eclipsing binaries detected in the OGLE SMC fields
and indicated a distinct period--luminosity--color relation.

Soszy{\'n}ski \etal (2004b) selected and studied a sample of 1660
ellipsoidal and eclipsing binaries with a red giant as one of the
components. They showed that the scatter of the {\it PL} relation strongly
depends on the amplitude of ellipsoidal variability -- the larger
amplitudes the tighter {\it PL} strip. It is understandable, because
variables with the largest amplitudes must lay close to the line defined in
the {\it PL} plane by systems with red giants filling up their Roche
lobes. Obviously this theoretical line indicates the short period edge of
the {\it PL} distribution, so the widening of the sequence must proceed
toward longer periods.

Consequently, the determination of the {\it PL} relation depends on
amplitudes of ellipsoidal variables -- larger amplitude variables have on
average shorter periods. To get consistent result in both galaxies, we cut
the sample of Soszy{\'n}ski \etal (2004b) at $A(I)=0.03$~mag. We excluded
also eclipsing variables, because eclipses of various depths randomly
change the mean magnitudes of the stars. Finally, Soszy{\'n}ski \etal
(2004b) showed that about 10\% of ellipsoidal red giants reveal clear
deformation caused by the eccentricity of their orbits. Periods of these
systems are systematically longer than periods of ellipsoidal variables
with circular orbits, therefore we removed the former group from our
sample. In the SMC we selected ellipsoidal red giants in the same manner as
in the LMC, and we found in total 440 objects.

Sequence~E stars are marked in yellow in Figs.~1--5. Note, that in Wood
\etal (1999) and in many subsequent papers this sequence was shifted toward
shorter periods by a factor of two, because half the orbital periods were
presented. The spread of sequence E in the NIR domain is considerably
larger than in the period--$W_I$ plane. It can be partly explained by
errors of the 2MASS photometry increasing quickly for fainter stars.

As it was noted in the previous Section, ellipsoidal variables in the LMC
follow at NIR passbands somewhat steeper {\it PL} relation than in the
SMC. Since the {\it PL} relation of ellipsoidal variables is a projection
of radius--luminosity dependence, this fact can be utilized for studying
parameters of red giant stars in environments of different metallicities.

\Section{Conclusions}
In this paper we showed the most complex structure of the {\it PL}
distribution presented so far. The Wood's five ridges turn out to be an
overlap of fourteen sequences (if consider closely spaced {\it PL}
relations of OSARGs, the number of sequences exceeds twenty). In order to
help recognizing {\it PL} relations with published {\it PL} laws we provide
Table~3 with appropriate identifications.

Diagrams employing period ratios (similar to the Petersen diagram) were
used as a tool for discriminating OGLE Small Amplitude Red Giants
(OSARGs). We compared three sequences of the {\it PL} relations for the LMC
OSARGs in the RGB phase with the calculated relation for first three radial
modes using isochrone calculations. We found an essential agreement with
our knowledge about metallicities and ages of red giants population in the
LMC. However, there are also discrepancies which may suggest the necessity
of refinement of the knowledge and/or stellar models. This is a potential
application of the OSARGs.

\MakeTableTop{l@{\hspace{4pt}}c@{\hspace{4pt}} l@{\hspace{8pt}} c@{\hspace{8pt}} c@{\hspace{8pt}} c@{\hspace{8pt}}}{12.5cm}
{Labels of the {\it PL} relations in this and previous papers}
{\hline
\noalign{\vskip3pt}
\multicolumn{3}{c}{this paper} & Wood \etal 1999 & Kiss and Bedding 2003 & Ita \etal 2004 \\
\noalign{\vskip3pt}
\hline
\noalign{\vskip3pt}
       &     & ${\rm b}_1$ &   & $R_1$ &                  \\
       & RGB & ${\rm b}_2$ & B & $R_2$ & $B^-$ \\
       &     & ${\rm b}_3$ & A & $R_3$ & $A^-$ \\
OSARGs &     & ${\rm a}_1$ &   &                           &                  \\
       &     & ${\rm a}_2$ & B & 2O                        & $B^+$ \\[-1ex]
       & \raisebox{1.5ex}{AGB} & ${\rm a}_3$ & A & 3O      & $A^+$ \\
       &     & ${\rm a}_4$ &   &                           &                  \\
\noalign{\vskip3pt}
\hline
\noalign{\vskip3pt}
         &  & C$_{\rm O}$ & C & F  & C \\[-1ex]
Miras    & \raisebox{1.5ex}{O-rich} & C$'_{\rm O}$ & B & 1O & C$'$ \\
and SRVs &  & C$_{\rm C}$ & C & F  & C \\[-1ex]
         & \raisebox{1.5ex}{C-rich} & C$'_{\rm C}$ & B & 1O & C$'$ \\
\noalign{\vskip3pt}
\hline
\noalign{\vskip3pt}
 & O-rich & D$_{\rm O}$ & D & $L_2$ & D \\[-1ex]
\raisebox{1.5ex}{LSPs} & C-rich & D$_{\rm C}$ & D & $L_2$ & D \\
\noalign{\vskip3pt}
\hline
\noalign{\vskip3pt}
\multicolumn{2}{l}{Ell. and Ecl.} & E & E$^*$ & $L_1^*$ & E$^*$ \\
\noalign{\vskip3pt}
\hline
\multicolumn{6}{l}{$^*$ -- the sequence is shifted due to halving the
orbital periods.}
}

Most likely mechanism responsible for pulsation in the OSARGs is the
stochastic excitation. Therefore we looked at these objects as solar-like
pulsators. The range of periods agrees with predictions based on
extrapolation of the relations found for much less luminous stars. The
amplitudes are lower than predicted by the Kjeldsen and Bedding (1995)
formula, where the the amplitude rises linearly with the luminosity-to-mass
ratio. However, there are calculations predicting a slower amplitude
rise. Testing theory of stochastic excitation is another possible
application of OSARGs.

We would like to bring the reader's attention to NIR Wesenheit index
($W_{JK}$). The sequences in the period--$W_{JK}$ plane are generally
better defined than those at $K_S$ magnitudes. Wesenheit index is a
reddening independent quantity, so even heavy reddened Miras fall on the
{\it PL} sequence. Therefore, the period--$W_{JK}$ relations can be used as
a distance indicators without correcting them for reddening. Moreover, 
O- and C-rich Miras and SRV obey very similar relations in the
period--$W_{JK}$ plane, while at $K_S$ the {\it PL} relations are
significantly different for both populations. We conclude that $W_{JK}$
index can be a useful tool for studying LPVs.

Comparison of the {\it PL} relations in NIR supports the binary star
scenario as the explanation of the LSP variability. We further develop this
scenario taking into account data on the ratios of the short period
(pulsational) and LSP variability. For OSARGs these ratio is nearly
constant which translates into nearly constant ratio of the stellar to
orbital radius the value of about 0.4. We proposed that the small mass
companion position is determined by the balance between mass loss and tidal
effects. This scenario requires that the proximity of the companion
enhances mass loss.  Moreover, it requires that a substantial fraction
(majority) of red giants have such companions and that their mass cannot be
too small.

\vspace*{9pt}

\Acknow{The paper was supported by the Foundation for Polish Science 
through the Homing Program and by MNiSW grants: 1P03D01130 and
N20303032/4275.

This publication makes use of data products from the Two Micron All Sky
Survey, which is a joint project of the University of Massachusetts and the
Infrared Processing and Analysis Center/California Institute of Technology,
funded by the National Aeronautics and Space Administration and the
National Science Foundation.}


\vspace*{9pt}

\begin{references}
\refitem{Bedding, T.~R.}{2003}{Astrophys. and Space Sci.}{284}{61}
\refitem{Bedding, T.~R., and Kjeldsen, H.}{2003}{Publ. Astron. Soc. Aust.}{20}{203}
\refitem{Christensen-Dalsgaard, J., Kjeldsen, H., and Mattei, J.A.}{2001}{\ApJ}{562}{141}
\refitem{Cioni, M.-R.L., Marquette, J.-B., Loup, C., Azzopardi, M., Habing,
H.J., Lasserre, T., and Lesquoy, E.}{2001}{\AA}{377}{945}
\refitem{Cioni, M.-R.L., \etal}{2003}{\AA}{406}{51}
\refitem{Clayton, M.L., and Feast, M.W.}{1969}{\MNRAS}{146}{411}
\refitem{Cutri, R.M., \etal}{2003}{~}{~}{``2MASS All-Sky Catalog of Point Sources''}
\refitem{Derekas, A., Kiss, L.L., Bedding, T.R., Kjeldsen, H., Lah, P., and
Szab{\'o}, G.M.}{2006}{\ApJL}{650}{L55}
\refitem{Dziembowski, W.A., Gough, D.O., Houdek, G., and Sienkiewicz, R.}{2001}{\MNRAS}{328}{601}
\refitem{Edmonds, P.D., and Gilliland, R.L.}{1996}{\ApJL}{464}{L157}
\refitem{Eggen, O.J.}{1977}{\ApJ}{213}{767}
\refitem{Feast, M.W., Glass, I.S., Whitelock, P.A., and Catchpole, R.M.}{1989}{\MNRAS}{241}{375}
\refitem{Fraser, O.J., Hawley, S.L., Cook, K.H., and Keller, S.C.}{2005}{\AJ}{129}{768}
\refitem{Gerasimovi\v{c}, B.P.}{1928}{Proceedings of the National Academy of Science}{14}{963}
\refitem{Glass, I.S., and Lloyd Evans, T.}{1981}{Nature}{291}{303}
\refitem{Groenewegen, M.A.T.}{2004}{\AA}{425}{595}
\refitem{Henry, G.W., Fekel, F.C., Henry, S.M., and Hall, D.S.}{2000}{\ApJS}{130}{201}
\refitem{Hinkle, K.H., Lebzelter, T., Joyce, R.R., and Fekel, F.C.}{2002}{\AJ}{123}{1002}
\refitem{Hughes, S.M.G., and Wood, P.R.}{1990}{\AJ}{99}{784}
\refitem{Ita, Y., Tanab{\'e}, T., \etal}{2004}{\MNRAS}{347}{720}
\refitem{Kholopov, P.N. \etal}{1985}{~}{~}{``General Catalog of Variable
Stars'', The Fourth Edition, Nauka, Moscow}
\refitem{Kiss, L.L., and Bedding, T.R.}{2003}{\MNRAS}{343}{L79}
\refitem{Kiss, L.L., and Bedding, T.R.}{2004}{\MNRAS}{347}{L83}
\refitem{Kjeldsen, H., and Bedding, T.R.}{1995}{\AA}{293}{87}
\refitem{Lebzelter, T., Schultheis, M., and Melchior, A.L.}{2002}{\AA}{393}{573}
\refitem{Madore, B.F.}{1982}{\ApJ}{253}{575}
\refitem{Noda, S., \etal}{2002}{\MNRAS}{330}{137}
\refitem{Osvalds, V., and Risley, A.~M.}{1961}{Publ. Leander McCormick Obs.}{11}{147}
\refitem{Pietrinferni, A., Cassisi, S., Salaris, M., and Castelli, F.}{2004}{\ApJ}{612}{168}
\refitem{Rucinski, S.M., and Maceroni, C.}{2001}{\AJ}{121}{254}
\refitem{Salaris, M., and Girardi, L.}{2005}{\MNRAS}{357}{669}
\refitem{Samadi, R., Goupil, M.-J., Alecian, E., Baudin, F., Georgobiani,
D., Trampedach, R., Stein, R., and Nordlund, A.}{2005}{Journal of
Astrophysics and Astronomy}{26}{171}
\refitem{Soszy\'nski, I., Udalski, A., Kubiak, M., Szyma\'nski, M.,
Pietrzy\'nski, G., \.Zebru\'n, K., Szewczyk,~O., and Wyrzykowski, {\L}.}{2004a}{\Acta}{54}{129}
\refitem{Soszy\'nski, I., Udalski, A., Kubiak, M., Szyma\'nski, M., 
Pietrzy\'nski, G., \.Zebru\'n, K., Szewczyk,~O., Wyrzykowski, {\L}., 
and Dziembowski, W.A.}{2004b}{\Acta}{54}{347}
\refitem{Soszy\'nski, I., Udalski, A., Kubiak, M., Szyma\'nski, M., 
Pietrzy\'nski, G., \.Zebru\'n, K., Szewczyk, O., Wyrzykowski, {\L}., and 
Ulaczyk, K.}{2005}{\Acta}{55}{331}
\refitem{Soszy\'nski, I.}{2007}{\ApJ}{660}{1486}
\refitem{Stello, D., et al.}{2007}{\MNRAS}{377}{584}
\refitem{Udalski, A.}{2003}{\Acta}{53}{291}
\refitem{Udalski, A., Kubiak, M., and Szyma\'nski, M.}{1997}{\Acta}{47}{319}
\refitem{Udalski, A., Szyma\'nski, M., Kubiak, M., Pietrzy\'nski, G., 
Soszy\'nski, I., Wo\'zniak, P., and \.Zebru\'n, K.}{1999}{\Acta}{49}{223}
\refitem{Udalski, A., Szyma\'nski, M., Kubiak, M., Pietrzy\'nski, G., 
Soszy\'nski, I., Wo\'zniak, P., and \.Zebru\'n, K.}{2000}{\Acta}{50}{307}
\refitem{Villaver, E., and Livio, M.}{2007}{\ApJ}{661}{1192}
\refitem{Wilson, R.E., and Merrill, P.W.}{1942}{\ApJ}{95}{248}
\refitem{Wood, P.~R.}{1995}{~}{~}{{\it IAU Colloq.~155}: ``Astrophysical 
Applications of Stellar Pulsation'', {\bf 83}, 127}
\refitem{Wood, P.R.}{2000}{Publ. Astron. Soc. Aust.}{17}{18}
\refitem{Wood, P.R., Olivier, E.A., and Kawaler, S.D.}{2004}{\ApJ}{604}{800}
\refitem{Wood, P.R., and Sebo, K.M.}{1996}{\MNRAS}{282}{958}
\refitem{Wood, P.R., \etal (MACHO team)}{1999}{~}{~}{in: {\it IAU Symp.}
{\bf 191}, ``Asymptotic Giant Branch Stars'', Ed. T.~Le~Bertre, A.~L\'ebre, 
and C.~Waelkens (San Francisco: ASP), 151}
\refitem{Zahn, J.-P.}{1977}{\AA}{57}{383}
\end{references}
\end{document}